\documentclass[12pt]{article}
\usepackage{amsfonts}

\newtheorem{lemma}{Lemma}
\newtheorem{remark}{Remark}
\begin{document}

\newtheorem{definition}{Definition}
\newtheorem{theorem}{Theorem}
\newtheorem{corollary}{Corollary}

\bigskip\bigskip

\centerline{\Large\bf On the physical meaning of the EPR--chameleon experiment}

\bigskip

\centerline{\sc Luigi Accardi\footnote{accardi@volterra.mat.uniroma2.it}, 
Kentaro Imafuku\footnote{imafuku@volterra.mat.uniroma2.it}, 
Massimo Regoli\footnote{massimo@volterra.mat.uniroma2.it}}\bigskip
\centerline{Centro Vito Volterra, Universit\`a degli Studi di
Roma ``Tor Vergata''}
\centerline{WEB-page: http://volterra.mat.uniroma2.it,}

\bigskip\bigskip\bigskip

\abstract{
The physical meaning of the EPR--chameleon 
experiment proposed in \cite{AcRe00b,AcRe01a},
in which the EPR correlations are reproduced by local,
independent, deterministic choices is re-examined. 
In addition we extend the mathematical model of \cite{AcRe00b,AcRe01a}
by showing that the dynamics considered there is effectively the reduced
dynamics of a fully reversible evolution.
We also propose a new protocol, more directly corresponding to
real experiments, in which the local computers only send back to the
central one the results of the evaluation of $\pm1$--valued functions. 
The program to run the experiment is available from the WEB-page: 
http://volterra.mat.uniroma2.it.}

\section{Physical interpretation of the experiment}\label{physint}

The present paper extends and clarifies the result of
the EPR--chameleon experiment, proposed in \cite{AcRe00b,AcRe01a}.
The goal of the experiment is (i) to construct a classical, deterministic,
reversible, dynamical system which reproduces the EPR correlations by local 
choices; (ii) to
use this model to give an experimental proof of the fact that macroscopic
systems with the above mentioned characteristics can violate Bell's
inequalities for principle and not for contingent reasons.

We do not pretend our model to be a hidden variable model for the EPR
experiments.

Our goal is to prove that Bell's statement \cite{Be64}: ``... the
statistical predictions of quantum mechanics are incompatible with local
predetermination ...'' is theoretically and experimentally unjustified
if by "statistical predictions of quantum mechanics" we mean the EPR
correlations and by "local predetermination" the possibility of
reproducing these correlations by a classical deterministic macroscopic
system subject to local choices.

The main hidden mathematical assumption in Bell's argument was pointed out
in \cite{Ac81a} (cf. \cite{AcRe00b} for a survey) where the nonkolmogorovian
character of the EPR correlations was first noticed and the crucial role
played by conditional probability emphasized.

In \cite{Ac93} (cf. also \cite{AcRe00b,AcRe01a})
von Neumann's measurement theory was extended to include in it the
two basic conditions of locality and causality and the physical
principle which allows to exploit Bell's hidden mathematical assumption
was individuated in the ``chameleon effect'': the dynamics of a system
may depend on the observables we want to measure (or, more generally, on
the local environment).

For such systems what you measure
is a {\it response to an interaction\/} and therefore, when dealing with them,
one should not speak of platonic (i.e. in principle unobservable) things such
as "values of non measured observables", but rather of ''instruction kits
telling them what to do when meeting a measurement apparatus'' \cite{[Tar98]}.
This is opposed to ballot box (or Einstein) reality in which you measure
what was there (independently of the environment).

More recently the main idea of the chameleon effect, i.e. that the local
interaction with the apparatus may have observable effects on the global
statistics, has been used to construct models related to the ``detection
loophole'' \cite{[GiGi01],[FiSZ99],[BrClTa99],[Ste99]} (cf. \cite{[Pea70],[CH74],[CHSH],[LoSh81],[GaMe87]} for earlier discussions).
In these papers one constructs local hidden variable models in which,
by assuming a partial inefficiency of the detectors, one can riproduce the
EPR correlations (or similar).

On the contrary, the dynamical theory of \cite{Ac93} is based on first
principles, and does not introduce contingent limitations of efficiency:
even postulating 100\% efficiency of all the instruments involved, because
of the chameleon effect, the local deterministic dynamics of the two particles
are different and one can use this freedom to construct examples of local
deterministic classical dynamical systems as specified by:

-- a state space (hidden parameters)

-- a deterministic dynamics (chameleon effect)

-- an initial probability measure (preparation of the experiment)

whose statistics is non classical.

The present paper produces a concrete example of such a system, which
reproduces the EPR correlations hence violates Bell's inequality. Moreover we give
an experimental proof of its local simulability.

The conceptual and experimental differences between the present approach
and the detection loophole is discussed in Section (\ref{expdif}) below.
Beyond the physical considerations, discussed in this section, there is also
a mathematical argument that proves the disjoint nature of the two topics
concerning coincidences and efficiency of the detectors. In all papers
involving efficiencies of detectors, these are explicitly built into the
model. On the contrary, as anybody can verify looking at Section
(\ref{detmod}) below, {\bf in our mathematical model neither inefficiencies
nor coincidences play any role. We have only a deterministic dynamics and
an initial probability distribution}. It is only at the simulation level
that the natural interpretation of the local factors as conditional
probabilities over coincidences emerges.
This is because a factor $2\pi$, appearing in a change of variables both 
in numerator and denominator is simplified in the calculations but 
{\bf must not} be simplified in order to allow local simulation 
(cf. Section (2), (3), (4)).

Our experiment describes the following classical dynamical system.
A source $C$ (central computer) produces pairs of particles $(S_1,S_2)$
which travel in different directions and after some time, each particle
interacts with a measurement apparatus $(M_1,M_2)$. By the {\it chameleon
effect\/} the dynamical evolution of each particle depends on the 
setting of the nearby apparatus, but not on the setting of the apparatus 
interacting
with the other particle (locality).

Even if, as just stated, our model has not the pretense to mimick the
real singlet experiments, an analogy with them will be useful for the
intuition of what is going on in it.
In this analogy one can interpret $(S_1,S_2)$ as a pair of photons and
$M_1$, $M_2$ as polarizers. The detector and the coincidence counters as
well as the space--time trajectories of the two particles
are not explicitly modeled in our dynamical system.

Following the standard prescriptions of (von Neumann) measurement theory, we
model the joint evolution of the system $S=(S_1,S_2)$ (i.e. the two particles)
and the apparatus $M=(M_1,M_2)$ (i.e. the two measurement apparata).
We want to incorporate in von Neumann's scheme of measurement the two
requirements of locality and causality. The local structure of our
dynamical system will be reflected in the model through:
\begin{description}
\item[(i)] a local structure of the initial state (probability measure) of
the composite $(S,M)$ system (cf. (\ref{locmeas}))
\item[(ii)] a local structure of the dynamics of the sub--systems
$(S_1,M_1)$ (particle 1, apparatus 1), $(S_2,M_2)$ (cf. (\ref{dyn}),
(\ref{dyn1}), (\ref{dyn2})).
\end{description}

We formulate these locality conditions only in the classical case and
for our specific model, but there is no difficulty in rephrasing them so
to include the general (classical and quantum) case.

We assume that, at the moment of emission from the source, the two particles
are in the same (microscopic) state (cf. (\ref{measys})) and that the state
of the apparatus is not changed by the interaction with the particle. For
example, if the polarizer was oriented in direction $a$ before interacting
with the photon, the same will be true after interaction (cf. (\ref{measa1}),
(\ref{measa2})).

The causality condition is reflected in the fact that the initial state
of $(S_1,S_2)$ does not depend on the settings of the apparata $M_1$,
$M_2$: in fact, at the time of emisison from the source, the particles
cannot know which measurements will be done on them.

We assume moreover that the dynamics of the two sub--systems $(S_1,M_1)$,
$(S_2,M_2)$ are independent, i.e. either sub--system does not feel the
influence of the other one. This is the locality condition as formulated
by EPR, Bell,... . We discretize time and consider only the initial and
final time of the experiment.

To each particle $S_1$, $S_2$ we associate a set of observables
$$\{S^{(1)}_a:a\in[0,2\pi\}\}\ ;\qquad\{S^{(2)}_b:b\in[0,2\pi]\}$$
These observables are modeled by functions defined on a space $\Omega$
and with values $+1$ or $-1$.

In our model just as for
photons, the actual detection takes place after interaction with the
polarizers. Thus, if experimenters $1$, $2$ want to measure $S^{(1)}_a$,
$S^{(2)}_b$ respectively then they will prepare the polarizer (or
magnet) $1$, $2$ oriented in direction $a$ (resp. $b$). In other words the
initial state of the polarizer depends on the observable we want to
measure. Since the particle interacts with the polarizer, the same will
be true for its dynamics. In other words: the chameleon effect is a
natural consequence of standard measurement theory.

Another consequence of what just said is that the state of the composite
system $(S_1,S_2$, $M_1,M_2)$ will depend on the pair of measurements $a,b$.
This is the known contextuality requirement but, as shown in section
(\ref{context}), by itself this is by far not sufficient to rule out the
validity of Bell's inequality: for this a more subtle analysis is
required.

Because of the different dynamics (chameleon effect) the two
particles have different trajectories hence even if they leave the
source at the same time, they do not necessarily interact
simultaneously with the corresponding apparatus. When they do we say that a
coincidence takes place. In agreement with what is done
in all the EPR type experiments also in our experiment the statistical 
countings are 
conditioned on coincidences because the correlations are equal time
correlations: this also assures that only photons coming
from entangled pairs are considered in the correlations. Since the
measurement apparata are approximatively equidistant from the source,
if the time interval between the emission of two consecutive pairs is
much larger than the coincidence interval (i.e. the time interval
within which two events are considered as simultaneous) then, assuming
100\% efficiency of the detectors and of the clocks (what we will
always do in the present
paper), we can be sure that only particles belonging to the
same pair can give rise to coincidences.

In EPR type experiments coincidences are measured either with very
precise clocks (coincidence intervals of order of nanoseconds) as in
\cite{[WJSWZ98]} or by direct connection of the polarizers to a coincidence
counter, as in \cite{SAWJST99}. The former technique allows a larger space
separation among the polarizers (a feature which is relevant for quantum
cryptography); the latter, as advocated in \cite{SAWJST99},
improves precision in the coincidence counting.

If the initial state of the two particles is chosen at
random, also the
number of coincidences will be random. Assuming reasonable ergodic
properties of the system, we can expect that this number will have small
fluctuations around its mean value.

In our model this mean value is independent of any special choice, in
particular it does not depend on the setting of the far away apparata.
Also this fact is in agreement with the experimental fact that: {\it
the size of the selected sample is found constant\/} \cite{AsGrRo82}.

\section{A local, deterministic, reversible, classical dynamical system
realizing the EPR correlation}\label{detmod}

In this section we describe the mathematical model on which the computer
experiment is based. According to the general description of the
chameleon effect \cite{AcRe00b,AcRe01a} we need an initial probability
measure and a dynamics. We construct these objects in the present
section and we will discuss their interpretation in the following
sections.

Define, for $ \sigma_1 ,\sigma_1 \in [0,2\pi]$, the functions
\begin{equation}
T'_{1,a}(\sigma_1)=\frac{\sqrt{2\pi}}{4}|\cos(\sigma_1-a)|,\quad
T'_{2,b}(\sigma_2)=\sqrt{2\pi}\label{1}
\end{equation}
and define the maps (dynamics)
\begin{equation}\label{dyn}
T_{1,a}(p_1)=(s_{1,a}(p_1),m_{1,a}(p_1))\quad ,\quad
T_{2,b}(p_2)=(s_{2,b}(p_2),m_{2,b}(p_2))
\end{equation}
\begin{equation}\label{dyn1}
p_1=(\sigma_1,\lambda_1)\ ;\quad s_{1,a}(\sigma_1,\lambda_1)=\sigma_1,\quad
m_{1,a}(\sigma_1,\lambda_1)=\lambda_1\frac{1}{T'_{1,a}(\sigma_1)}
\end{equation}
\begin{equation}\label{dyn2}
p_2=(\sigma_2,\lambda_2)\ ;\quad s_{2,b}(\sigma_2,\lambda_2)=\sigma_2,\quad
m_{2,b}(\sigma_2,\lambda_2)=\lambda_2\frac{1}{T'_{2,b}(\sigma_2)}
\end{equation}
Define moreover the measures $(\sigma_1,\lambda_1, \sigma_2,\lambda_2
\in [0,2\pi])$:
\begin{equation}\label{measys}
p_{S}(\sigma_1,\sigma_2)=\frac{1}{2\pi}\delta(\sigma_1-\sigma_2)
d\sigma_1d\sigma_2
\end{equation}
\begin{equation}\label{measa1}
p_{1,a}(\sigma_1,\lambda_1)=\delta(m_{1,a}(\sigma_1,\lambda_1)-m_a)d\lambda_1
\end{equation}
\begin{equation}\label{measa2}
p_{2,b}(\sigma_2,\lambda_2)=\delta(m_{2,b}(\sigma_2,\lambda_2)-m_b)d\lambda_2
\end{equation}

\begin{lemma}\label{normlm} The measure
\begin{equation}\label{locmeas}
p_{S}(\sigma_1,\sigma_2)p_{1,a}(\sigma_1,\lambda_1)
p_{2,b}(\sigma_2,\lambda_2)
d\sigma_1\sigma_2d\lambda_1d\lambda_2
\end{equation}
is a probability measure on $[0,2\pi]^4$.
\end{lemma}

\noindent{\it Proof\/}. The positivity is obvious. The normalization
condition
\begin{equation}
\int ~
p_{S}(\sigma_1,\sigma_2)p_{1,a}(\sigma_2,\lambda_1)
p_{2,b}(\sigma_2,\lambda_2)
d\sigma_1\sigma_2d\lambda_1d\lambda_2=1
\end{equation}
follows from a simple calculation (cf. also the proof of Lemma
(\ref{eprlm})).

\begin{remark} It would be tempting to interpret the measures
(\ref{measa1}), (\ref{measa2}) as conditional distributions of the state of
the apparatus given the incoming state of the correponding photon. However this
is not possible because it happens that:
\begin{equation}\label{nonorm}
\int p_{1,a}(\sigma_1,\lambda_1) d\lambda_1 =
{\sqrt{2\pi}\over4}\,|\cos(\sigma_1-a)|
\neq 1,\quad
\int p_{2,b}(\sigma_2,\lambda_2) d\lambda_2=\sqrt{2\pi}\neq 1
\end{equation}
As shown in section (\ref{context}) the inequalities in (\ref{nonorm}) are
necessary for the violation of Bell's inequality.
\end{remark}

\begin{lemma}\label{eprlm} Let the $\pm1$--valued
maps $S^{(1)}_{a}(\sigma_1,\mu_1), S^{(2)}_{x}(\sigma_2,\mu_2)$
$(\sigma_j,\mu_j\in[0,2\pi])$ be given by
\begin{equation}\label{choice}
S^{(1)}_{a}(\sigma,\mu)={\mbox{sgn}}(\cos(\sigma-a))  \quad ,\quad
S^{(2)}_{x}=-S^{(1)}_{x}
\end{equation}
then, in the above notations,
\begin{eqnarray}
&  &\int~{S_{1}^{(a)}
\left(s_1(\sigma_1,\lambda_1),m_1(\sigma_1,\lambda_1)\right)}
{S_{2}^{(b)}\left(s_2(\sigma_2,\lambda_2),m_2(\sigma_2,\lambda_2)\right)} 
\label{apo1} \\
&&
\times p_{S}(\sigma_1,\sigma_2)
p_{1,a}(\sigma_1,\lambda_1)p_{2,b}(\sigma_2,\lambda_2)
d\sigma_1 d\sigma_2 d\lambda_1 d\lambda_2 =-\cos(a-b)
\nonumber
\end{eqnarray}
\end{lemma}

\noindent{\it Proof\/}. With the choices (\ref{measys}), (\ref{measa1}),
(\ref{measa2}), the correlations (\ref{apo1}) become
\begin{equation}
\int\int\int\int S^{(1)}_a(s_{1,a}(\sigma_1,\lambda_1),m_{1,a}(\sigma_1,
\lambda_1))S^{(2)}_b(s_{2,b}(\sigma_2,\lambda_2),m_{2,b}(\sigma_2,\lambda_2))
\label{apo2}
\end{equation}
$$\delta(m_{1,a}(\sigma_1,\lambda_1)-m_a)\delta(m_{2,b}(\sigma_2,
\lambda_2)-m_b)p_{S}(\sigma_1,\sigma_2)d\lambda_1d\lambda_2d\sigma_1d
\sigma_2$$

Changing variables
$$m_{1,a}(\sigma_1,\lambda_1)=\mu_1$$
$$m_{2,b}(\sigma_2,\lambda_2)=\mu_2$$
$$m'_{1,a}(\sigma_1,\lambda_1)d\lambda_1=d\mu_1$$
$$m'_{2,b}(\sigma_2,\lambda_2)d\lambda_2=d\mu_2$$
and noting that for almost all $a,b,\sigma_1,\sigma_2\in [0,2\pi]$ the
functions
\begin{equation}
m_{1,a}(\sigma_1,\cdot),~m_{2,b}(\sigma_2,\cdot):
[0,2\pi]\rightarrow [0,2\pi]
\end{equation}
are invertible, one has from (\ref{1}), (\ref{dyn1}), (\ref{dyn2})
\begin{equation}
d\lambda_1={1\over m'_{1,a}(\sigma_1,m^{-1}_{1,a}(\sigma_1,\mu_1))}\,
d\mu_1=:\tilde T'_{1,a}(\sigma_1,\mu_1) d\mu_1
\label{jac1}
\end{equation}
\begin{equation}
d\lambda_2={1\over m'_{2,b}(\sigma_2,m^{-1}_{2,b}(\sigma_2,\mu_2))}\,
d\mu_2=:\tilde T'_{2,b}(\sigma_2,\mu_2)d\mu_2
\label{jac2}
\end{equation}
and, after the change of variables, (\ref{apo2}) becomes
$$\int\int\int\int S^{(1)}_a(s_{1,a}(\sigma_1,m^{-1}_{1,a}(\sigma_1,
\mu_1)),\mu_1)S^{(2)}_b(s_{2,b}(\sigma_2,m^{-1}_{2,b}(\sigma_2,\mu_2)),\mu_2)$$
\begin{equation}
\tilde T'_{1,a}(\sigma_1,\mu_1)\tilde T'_{2,b}(\sigma_2,\mu_2)
\delta(\mu_1-m_a)\delta(\mu_2-m_b)p_{S}(\sigma_1,\sigma_2)d\mu_1
d\mu_2d\sigma_1d\sigma_2\label{apo3}
\end{equation}
Because of our choice of the functions $S^{(1)}_a, S^{(2)}_b, T'_{1,a},
T'_{2,b}$, these have the form
$$S^{(1)}_a(\sigma_1):=S^{(1)}_a(s_{1,a}(\sigma_1,
m^{-1}_{1,a}(\sigma_1,m_a),\mu_1)$$
$$S^{(2)}_b(\sigma_2):=S^{(2)}_b(s_{2,b}(\sigma_2,
m^{-1}_{2,b}(\sigma_2,m_b),\mu_2)
$$
\begin{equation}
T'_{1,a}(\sigma_1):=\tilde T'_{1,a}(\sigma_1,m_a)\label{jac3}
\end{equation}
\begin{equation}
T'_{2,b}(\sigma_2):=\tilde T'_{2,b}(\sigma_2,m_b)\label{jac4}
\end{equation}
in the sense that the right hand side depends only on the variables
written on the left hand side. Therefore (\ref{apo3}) becomes
\begin{equation}
\int\int S^{(1)}_a(\sigma_1)S^{(2)}_b(\sigma_2)T'_{1,a}(\sigma_1)
T'_{1,b}(\sigma_2)p_{S}(\sigma_1,\sigma_2)d\sigma_1d\sigma_2\label{19a}
\end{equation}
Finally, since
\begin{equation}
p_{S}(\sigma_1,\sigma_2)={1\over 2 \pi}\delta(\sigma_1-\sigma_2)\label{19b}
\end{equation}
we arrive at
$$\int S^{(1)}_a(\sigma)S^{(2)}_b(\sigma)T'_{1,a}(\sigma)T'_{1,b}
(\sigma) {d\sigma \over 2 \pi}=-\int_0^{2\pi} \cos(\sigma-b)\hbox{sgn}
(\cos(\sigma-a)) d\sigma =-\cos(b-a)$$
which is the thesis.

\begin{remark} Given (\ref{1}) the expression (\ref{19a}), (\ref{19b})
is precisely the one which was taken as starting point in the paper
\cite{AcRe01a}. The result of the present section proves that the
interpretation in terms of reduced dynamics proposed in that paper was
correct: the Heisenberg dynamics used in \cite{AcRe01a} is indeed the
reduced dynamics of the reversible dynamics considered in the present
paper.
\end{remark}

\section{The simulation problem}\label{simprob}

Starting from this section until the end of Section (\ref{sec_cond})
we describe two methods of local simulation of the
deterministic model described in Section (\ref{detmod}). The method we
actually used in the realization of our experiment, is the one described
in Section (\ref{sec_cond}). The other method corresponds to
previous versions of the experiment \cite{AcRe01a}
(this method corresponds to the
option ``old'' in the menu of the programme in the web)
and is interesting because it illustrates the
difference between the theoretical problem, solved in Section 2, and the
actual simulation method.

In order to simulate locally the probability measure (\ref{locmeas}) we
normalize the local factors at $a$ and $b$, obtaining
the following expression for the correlations (\ref{apo1})
\begin{equation}
\int^{2\pi}_0\int^{2\pi}_0{d\sigma_1d\sigma_2\over2\pi}\,\delta
(\sigma_1-\sigma_2){2\pi\over4}\,|\cos(\sigma_1-a)|\label{19c}
\end{equation}
$$\left[\int^{2\pi}_0 S^{(1)}_a(\sigma_1,\lambda_1)
{p_{1,a}(\sigma_1,\lambda_1)d\lambda_1\over Z^{(1)}_a(\sigma_1)}\right]
\left[\int^{2\pi}_0 S^{(2)}_b(\sigma_2,\lambda_2)
{p_{2,b}(\sigma_2,\lambda_2)d\lambda_2\over Z^{(2)}_b(\sigma_2)}\right]$$

Integrating the $\delta$--function, simplifying the $2\pi$--factor and
introducing the notations
\begin{equation}
\tilde p_{j,x}(\sigma_j,\lambda_j)d\lambda_j={p_{j,x}
(\sigma_j,\lambda_j)\over Z^{(j)}_x(\sigma_j)}\,d\lambda_j\quad;\qquad
j=1,2\ ;\quad x=a,b\label{19d}
\end{equation}
($Z^{(j)}_a(\sigma_j)$ being the normalization factor)
we obtain
$$\int^{2\pi}_0d\sigma
\left[\int^{2\pi}_0{|\cos(\sigma-a)|\over4}\, S^{(1)}_a
(\sigma,\lambda_1)\tilde p_{1,a}(\sigma,\lambda_1)d\lambda_1\right]
\left[\int^{2\pi}_0 S^{(2)}_b(\sigma,\lambda_2)\tilde p_{2,b}(\sigma,
\lambda_2)d\lambda_2\right]$$

Given our choice (\ref{choice}) of $S^{(1)}_{a}$, $S^{(2)}_{b}$, the second integral 
in square brackets is equal to $S^{(2)}_b(\sigma)$.  This is evaluated locally 
by computer 2.  The integrand of the first integral in square brackets is equal to 
$
\frac{1}{4}\cos(\sigma-a).
$
In order to simulate locally this integral we replace in it the function 
$\frac{1}{4}|\cos(\sigma-a)| S^{(1)}_a(\sigma,\lambda_1)$ by any function 
$\hat{S}_a^{(1)}(\sigma,\lambda)$, with values $\pm 1$, such that 
$$
\int^{2\pi}_{0}\hat{S}_a^{(1)}(\sigma,\lambda)
\tilde{p}_{1,a}(\sigma,\lambda_1)d\lambda_1=\frac{1}{4}\cos(\sigma-a).
$$
Such function is easily constructed because 
$\tilde{p}_{1,a}(\sigma,\lambda_1)d\lambda_1$
is a probability measure. This integral is evaluated locally by computer 1.
After this, and with the notation
$$
\hat{S}_b^{(2)}(\sigma,\lambda)={S}_b^{(2)}(\sigma,\lambda)
$$
we arrive to integral 
$$\int^{2\pi}_0d\sigma
\left[\int^{2\pi}_0\hat S^{(1)}_a(\sigma,\lambda_1)\tilde p_{1,b}(\sigma,
\lambda_1)d\lambda_1\right]
\left[\int^{2\pi}_0\hat S^{(2)}_b(\sigma,\lambda_2)\tilde p_{2,b}(\sigma,
\lambda_2)d\lambda_2\right]$$
\begin{equation}
=\int^{2\pi}_0d\sigma{\cos(\sigma-a)\over4}\,(-\hbox{sgn}(\cos(\sigma-
b)))\label{startp}
\end{equation}
To simulate this integral directly would be easy (this is what was done in the 
first experiment \cite{AcRe01a}).
However this would require that the integrals in 
$d\lambda_1$, $d\lambda_2$ should be calculated locally by computer 1 and 2 
respectively and only the two integrals in square brackets would be sent to the central computer.  If, in greater coherence with the experiments, we want 
the local computers to send to the central computer only sequences of $\pm1$ 
(i.e., the values of $\hat S_a^{(1)}(\sigma,\lambda_1), 
\hat S_b^{(2)}(\sigma,\lambda_2)$), then we have to generate a sequence of 
$(\sigma_j)$ in $[0,2\pi]$ which realizes this goal.  To this purpose, 
following the standard Monte Carlo procedure, we normalize the 
$d\sigma$-integral
by dividing and multiplying by $2\pi$
\begin{equation}
2\pi\left[\int^{2\pi}_0{d\sigma\over2\pi}\,{\cos(\sigma-a)\over4}\,
(-\hbox{sgn}(\cos(\sigma-b)))\right]\label{startp2}
\end{equation}
this leads to the integral 
\begin{equation}
2\pi\left\{\int^{2\pi}_0{d\sigma\over2\pi}\,\left[\int^{2\pi}_0 
\hat S^{(1)}_a(\sigma,\lambda_1)
\tilde p_{1,a}(\sigma,\lambda_1)d\lambda_1\right]\left[\int^{2\pi}_0 
\hat S^{(2)}_b(\sigma,\lambda_2)
\tilde p_{2,b}(\sigma,\lambda_2)d\lambda_2\right]\right\}\label{start3p}
\end{equation}
in which one can now easily approximate locally each piece with Riemann sums 
\begin{equation}
{2\pi\over N}\,\sum_{\sigma_j}\left[{1\over K_1}\,\sum_{k_1}\hat S^{(1)}_a
(\sigma_j,\lambda^{(j)}_{a,k_1})\right]\left[{1\over K_2}\,\sum_{k_2}
\hat S^{(2)}_b(\sigma_j,\lambda^{(j)}_{b,k_2})\right]\label{26a}
\end{equation}
This gives the following prescriptions: let
\begin{description}
\item[$C$] denote central computer
\item[$A$] Experimenter 1
\item[$B$] Experimenter 2
\end{description}

\begin{enumerate}
\item $C$ produces any sequence $\sigma_j$ in $[0,2\pi]$ distributed according to
${d\sigma\over2\pi}$. It is not necessary that the sequence $(\sigma_j)$
has good chaotic properties: any equidistributed sequence in $[0,2\pi]$
can be used.
\item For each $\sigma_j$, $A$ produces $(\lambda^{(j)}_{a,k_1})_{k_1}$
distributed according to $\hat p_{1,a}(d\lambda)$ and produces the sequence of
values $\hat S^{(1)}_a(\sigma_j,\lambda^{(j)}_{a,k_1})\in\{\pm1\}$
\item For each $\sigma_j$, $B$ produces $(\lambda^{(j)}_{b,k_2})_{k_2}$
distributed according to $\hat p_{2,b}(d\lambda)$ and produces the sequence
of values $\hat S^{(2)}_b(\sigma_j,\lambda^{(j)}_{b,k_2})\in\{\pm1\}$.
\item $A$ and $B$ send these two sequences to $C$
\item $C$ takes the arithmetic mean of the product of the two sequences
and multiplies the result by $2\pi$ thus compensating the division by
$2\pi$, introduced in Step (1).
\end{enumerate}

\begin{remark} The above simulation procedure has two unsatisfactory
features: (i) the artificial (although standard in Monte Carlo
simulations) ``multiplication and division by $2\pi$'';
(ii) the fact that, before multiplication by $2\pi$, the correlations
(\ref{26a}) for $a=b$ are not equal to $-1$ but to $-1/2\pi$, hence the
singlet condition $S^{(1)}_a=-S^{(2)}_b$, although globally satisfied
due to Lemma (2), is not satisfied at each step of the simulation
procedure.
\end{remark}

In Section (5) below we describe a simulation method which is free from
these drawbacks. Before that, in the following section, we describe the
probabilistic origins of this difficulty.

\section{Probabilistic interpretation of the multiplication and division 
by $2\pi$ \label{imaima}}

In this section we propose a probablistic interpretation of the
decomposition (\ref{19c}) before simplification of the two $2\pi$
factors. This interpretation gives the key for the
simulation used in Section (\ref{sec_cond}).

Let us first discuss, in some generality, which kind of properties it is
reasonable to expect from a ``locally simulable'' measure, i.e. a
measure that, in addition of having the local structure (\ref{locmeas})
is also such that it can be split into 3 pieces each of which is
simulable in a different computer.

Recall that in real experiments the countings are conditioned on coincidences
and that these occur at the polarizers site. By locality we can expect,
for a locally simulable probability measure a structure of the form
$$p_S(\sigma_1,\sigma_2)E_{\Gamma_c}(\cdot)(\sigma_1,\sigma_2)d\sigma_1
d\sigma_2$$
where $p_S$ is the distribution at the source and
$E_{\Gamma_c}(\cdot)(\sigma_1,\sigma_2)$ denotes the conditional
distribution of the apparatus variables given simultaneous interaction
with particles $S_1,S_2$ in the state $\sigma_1,\sigma_2$ respectively. 
Denoting $\Gamma_c$ the event: ``a coincidence occurs at the time of
measurement'', the conditional expectation $E_{\Gamma_c}$ can be written
as
$$E_{\Gamma_c}(F)={E(F\chi_{\Gamma_c})\over P(\Gamma_c)}$$
where $F$ is any function, $\chi_{\Gamma_c}(\omega)=1$ if 
$\omega\in\Gamma_c$ and $=0$ if $\omega\notin\Gamma_c$ and 
$P(\Gamma_c)$ is the probability of $\Gamma_c$.

Moreover it is reasonable to expect that, given coincidence and the pair
$(\sigma_1,\sigma_2)$, the results of measurement at instruments 1 and 2
should be independent events: in fact, once given $\Gamma_c$,
$\sigma_1$, $\sigma_2$, what happens at 1 (resp. 2) only depends on the
local apparatus variable $\lambda_1$ (resp. $\lambda_2$). Thus, if
$F_{1,a}$, $F_{2,b}$ are local observables (local in the sense that
$F_{j,x}(\sigma_1,\sigma_2,\lambda_1,\lambda_2)=F_{j,x}(\sigma_j,
\lambda_j)$ $(j=1,2,x=a,b)$), one should have
$$E(F_{1,a}F_{2,b}\chi_{\Gamma_c})(\sigma_1,\sigma_2)=E(F_{1,a}
\chi_{\Gamma_c})(\sigma_1,\sigma_2)E(F_{2,b}\chi_{\Gamma_c})(\sigma_1,
\sigma_2)$$
Our model is Section (2) also satisfies the additional condition
$$E(F_{j,x}\chi_{\Gamma_c})(\sigma_1,\sigma_2)=E(F_{j,x}\chi_{\Gamma_c})
(\sigma_j)\ ;\quad j=1,2$$
but a model not satisfying this condition would not violate locality
because $(\sigma_1,\sigma_2)$ is the state at the source, where the 2
particles may be very near each other.

In any case, assuming both the above conditions,
we would find for the correlations
$$\langle S^{(1)}_aS^{(2)}_b\rangle=\int\int p_S(\sigma_1,\sigma_2)
d\sigma_1d\sigma_2E_{\Gamma_c}(S^{(1)}_aS^{(2)}_b)(\sigma_1,\sigma_2)$$
\begin{equation}
={1\over P(\Gamma_c)}
\int\int p_S(\sigma_1,\sigma_2)d\sigma_1d\sigma_2E_{1,a}(S^{(1)}_a
\chi_{\Gamma_c})(\sigma_1)E_{2,b}(S^{(2)}_b
\chi_{\Gamma_c})(\sigma_2)\label{condcor}
\end{equation}
Now if, as it is done in many imprecise probabilistic models of the EPR
experiment, one neglects the conditioning factor $1/P(\Gamma_c)$, one
would be lead to consider, instead of the correct integral
(\ref{condcor}), the integral
\begin{equation}
\int\int p_S(\sigma_1,\sigma_2)d\sigma_1d\sigma_2E_{1,a}(\hat S^{(1)}_a)
(\sigma_1)E_{2,b}(\hat S^{(2)}_b)(\sigma_2)\label{nccor}
\end{equation}
$(\hat S^{(j)}_x=S^{(j)}_x\chi_{\Gamma_c})$. To this integral one can
easily apply the considerations of Section (7) below and deduce the Bell
inequality
\begin{equation}
|\langle S^{(1)}_aS^{(2)}_b\rangle_0-\langle S^{(1)}_cS^{(2)}_b\rangle_0|
-\langle S^{(1)}_aS^{(2)}_c\rangle_0\leq1\label{incori}
\end{equation}
where $\langle\cdot\rangle_0$ means that the correlations are computed
with the incorrect integral (\ref{nccor}). However, if the correlations
are computed with the correct integral (\ref{condcor}) one has to
multiply both sides of (\ref{incori}) by $1/P(\Gamma_c)$ which leads to
\begin{equation}
|\langle S^{(1)}_aS^{(2)}_b\rangle-\langle S^{(1)}_cS^{(2)}_b\rangle|
-\langle S^{(1)}_aS^{(2)}_b\rangle\leq{1\over P(\Gamma_c)}\label{cori}
\end{equation}

Now, in the notation (\ref{19d}), if we specify our system by the
requirements that $p_S(\sigma_1,\sigma_2)d\sigma_1d\sigma_2$ is given by
(\ref{measys}) and:
$$P(\Gamma_c)=(2\pi)^{-1}$$
$$E_{1,a}(S^{(1)}_a\chi_{\Gamma_c})(\sigma_1)=\int^{2\pi}_0{|\cos(\sigma_1-
a)|\over4}\,S^{(1)}_a(\sigma_1,\lambda_1)\tilde p_{1,a}(\sigma_1,
\lambda_1)d\lambda_1$$
$$E_{2,b}(S^{(2)}_b\chi_{\Gamma_c})(\sigma_2)=\int^{2\pi}_0S^{(2)}_b
(\sigma_2,\lambda_2)\tilde p_{2,b}(\sigma_2,\lambda_2)d
\lambda_2$$
then we find the expression (\ref{19c}).

The choice we made in our experiment is very special. Formula
(\ref{cori}) suggests that, by appropriately constructing deterministic
dynamical systems, one can make the probability $P(\Gamma_c)$ arbitrarily
small, hence the bound in (\ref{cori}) arbitrarily high.

\section{Conditioning on coincidences: direct simulation \label{sec_cond}}

In the present section we describe a technique to simulate directly the
conditional probabilities introduced in section (\ref{imaima}). In this way the
multiplication and division by $2\pi$ comes out from the statistical
countings themselves and the singlet condition is verified at each step
of the simulation.

In the idealized dynamical system considered in our experiment we
consider only two time instants 0 (initial) and 1 (final) so, in our
case, a ``trajectory'' consists of a single jump. We do not describe the
space--time details of the trajectory because we are only interested in
distinguishing 2 cases:
\begin{description}
\item[--] at time 1 the particle is in the apparatus (and in this case
it is detected with certainty)
\item[--] at time 1 the particle is not in the apparatus (and in this
case it makes no sense to speak of detection)
\end{description}
Thus our ``configuration space'' for the single particle will be made of
3 points: $s$ (source), 1 (inside apparatus), 0 (outside apparatus).
Since at time 0 the ``position'' of both particles is always $s$,
because of the chameleon effect, the position $q_{j,1}$ of particle
$j=(1,2)$ at time 1 will depend on the polarization $a_j$, on the
initial state $\sigma$ and on the state $\lambda_j$ of the apparatus
$M_j(j=1,2)$:
$$q_{j,1}=q_{j,1}(a_j,\sigma,\lambda_j)\quad;\qquad j=1,2,$$
The local, deterministic dynamical law of this dependence is the one
described in Section (5.4).

There is no conceptual difficulty to include in our
model the consideration of the space--time trajectory of
the particle. This surely would improve the present
model, however the main conclusion of our experiment, i.e. the
reproducibility of the EPR correlations by a classical, deterministic,
local dynamical system, will not change.

\subsection{Description of the experiment}

\begin{enumerate}
\item Let $N \leq N_{tot}$ be natural integers and let
\begin{equation}\label{seq}
\{\sigma_j \ : \ j = 1, \dots , N \}
\end{equation}
be the sequence of numbers
\begin{equation}
\{\sigma_j := (2\pi/N)\times j:j=1,\dots,N\}\label{sets}
\end{equation}
We have checked in several experiments that any pseudo--random sequence
in $[0,2\pi]$ with good equidistribution properties will lead to the
same result. This fact is reflected in the option $D$ (deterministic) or
$R$ (random) that has been inserted in the program of the experiment.
Let $N(\sigma_j)$ ($j = 1, \dots , N$) denote a sequence of natural integers
such that
$$\sum_{j=1}^{N} N(\sigma_j)=N_{tot}$$
\item For each $j$ from $1$ to $N$,
repeat the following 3 operations (a), (b), (c), $N(\sigma_j)$ times
\begin{enumerate}
\item The central computer sends $\sigma_j$ to the computers 1 and 2.
\item Computer 1 computes the position of particle 1 as described in section 
5.4 below and sends back $S_a^{(1)}(\sigma_j)$ ($=1$ or $-1$) if the particle 
is inside the apparatus. It sends back $\emptyset$ (empty) if the particle 
is outside the apparatus. Similarly Computer 2 does the same thing. 
The dynamics is such that $S^{(1)}_a(\sigma_j)$ is sent back with probability 
$p_{1,a}(\sigma_j)$ and $S^{(2)}_b(\sigma_j)$ is sent back 
with probability $p_{2,b}(\sigma_j)$ where
$p_{1,a},p_{2,b}$ are sufficiently regular
probability densities (say piecewise smooth with a finite number of
discontinuities in $[0,2\pi ]$ (See sec.\ref{howtorealise})).  
This corresponds in the real experiments, to labeling the local detection time 
of the photon. 
When both computers send back a value $\pm1$, then we say that a
{\it coincidence} occurs.
\item {\it Only in case of a coincidence, i.e. when the central computer
receives the value $\pm1$  from both computers}, the central computer computes
the ``correlation product'' $S_a^{(1)}(\sigma_j)S_b^{(2)}(\sigma_j)$.
\end{enumerate}
\item The central computer computes the correlation as
\begin{equation}\label{definition}
\frac{\mbox{Sum of all correlation products}}
{\mbox{The total number of coincidences}}.
\end{equation}
\end{enumerate}

\subsection{Computation of the correlations}

Introducing
\begin{equation}
p(\sigma_j)=\frac{N(\sigma_j)}{N_{tot}}
\end{equation}
the expected number of coincidences ${\cal N}_{\hbox{coincidences}}$ and
the sum of all correlation products ${\cal S}_{\hbox{correlations}}$ become
respectively
\begin{eqnarray}
{\cal N}_{\hbox{coincidences}}&=&
\sum_{j=1}^{N}N(\sigma_j)p_{1,a}(\sigma_j)p_{2,b}(\sigma_j)
=N_{tot}\sum_{j=1}^{N}p(\sigma_j)p_{1,a}(\sigma_j)p_{2,b}(\sigma_j)\\
{\cal S}_{\hbox{correlations}}& =&
\sum_{j=1}^{N}N(\sigma_j)p_{1,a}(\sigma_j)p_{2,b}(\sigma_j)
S^{(a)}_1(\sigma_j)S^{(b)}_2(\sigma_j)\nonumber\\
&=&N_{tot}\sum_{j=1}^{N}p(\sigma_j)p_{1,a}(\sigma_j)p_{2,b}(\sigma_j)
S^{(a)}_1(\sigma_j)S^{(b)}_2(\sigma_j)
\end{eqnarray}
Thus the correlation defined by (\ref{definition}) is
\begin{equation}
\frac{{\cal S}_{\hbox{correlations}}}{{\cal N}_{\hbox{coincidences}}} =
\frac{\sum_{j=1}^{N}p(\sigma_j)p_{1,a}(\sigma_j)p_{2,b}(\sigma_j)
S^{(a)}_1(\sigma_j)S^{(b)}_2(\sigma_j)}{\sum_{j=1}^{N}p(\sigma_j)p_{1,a}(\sigma_j)p_{2,b}(\sigma_j)}
\end{equation}
and therefore
\begin{equation}\label{integral}
\frac{{\cal S}_{\hbox{correlations}}}{{\cal N}_{\hbox{coincidences}}}
\rightarrow \frac{\int_{0}^{2\pi} d\sigma ~p(\sigma)p_{1,a}(\sigma)p_{2,b}(\sigma)
S^{(a)}_1(\sigma)S^{(b)}_2(\sigma)}{\int_{0}^{2\pi} d\sigma ~p(\sigma)
p_{1,a}(\sigma)p_{2,b}(\sigma)}
\end{equation}
where $p(\sigma)$ is a probability density with properties analogue to
$p_{1,a}$ and $p_{2,b}$.

\subsection{Realization of the EPR correlation}\label{epr}

With the choices:
\begin{equation}
p(\sigma)=\frac{1}{2\pi}\quad ,\quad
p_{1,a}(\sigma)=\frac{1}{4}|\cos(\sigma-a)|\quad ,\quad
p_{2,b}(\sigma)=1
\end{equation}
\begin{equation}
S_{1}^{(a)}(\sigma)=\hbox{sgn}(\cos(\sigma-a)),\quad
S_{2}^{(b)}(\sigma)=-\hbox{sgn}(\cos(\sigma-b))
\end{equation}
we obtain 
\begin{equation}\label{num}  
\hbox{numerator of (\ref{integral})} =  -\frac{1}{2\pi}\cos(a-b)
\end{equation}
\begin{equation}\label{denom}
\hbox{denominator of (\ref{integral})} = \frac{1}{2\pi}
\end{equation}
Therefore for large $N$ the correlation (\ref{definition}) is well
approximated by
\begin{equation}
-\cos(a-b)
\end{equation}
which is exactly the EPR correlation. We underline that, as shown by
(\ref{denom}), even if the
mechanism of coincidences depends on the setting of the apparatus, the
expected number of coincidences is independent of it, in agreement
with the experimental result quoted in the introduction.

\subsection{Computation of the position}\label{howtorealise}

With the choices of section (\ref{epr}), $p_{2,b}(\sigma)$ is trivial
$(=1)$ and this means that computer $2$ always sends back 
$S^{(b)}_2(\sigma)$ to the central computer.
On the other hand, computer $1$ associates the label 1 to $S^{(a)}_1(\sigma)$
in the following way,
\begin{enumerate}
\item For each input $\sigma_j$, 
generate a new random variable $\lambda_1$ with 
a probability distribution $P_{01}(\lambda_1)=\chi_{[0,1]}$.
\item When $\lambda_1$ is such that $0\le \lambda_1 \le p_{1,a}(\sigma)$
computer $1$ concludes that the particle is inside sends the value 
$S^{(a)}_{1}(\sigma_j)$, otherwise it sends $\emptyset$.
\end{enumerate}

\section{Difference between coincidences and efficiency of
the detectors}\label{expdif}

As already emphasized in the introduction of the present paper, the
difference between conditioning on coincidences and efficiency detectors
is a principle one. Moreover, in many experimental situations, the total
number of pairs emitted by the source is in principle unboservable. For
example in all EPR type experiments with photons, the source of
entangled pairs has a finite size hence the probability that one or both
photons of some pair is reabsorbed by the source itself is nonzero. In
all these cases the statistical counting is conditioned on the
simultaneously detected
pairs. Thus, whenever we want to estimate statistically equal time
correlations of the form
$$\langle S^{(1)}_a(t)S^{(2)}_b(t)\rangle$$
we must be aware that in general they will be correlations {\it
conditioned on coincidences\/}.

In the loophole argument, the
following fact has been noted: Einstein's local realism can be consistent
with the experimental data when the excessive correlations might be
possessed only by a fraction of the coincident pairs actually detected.
In other words, an high efficiency of the detectors is required to exclude
the loophole.

Conditioning on coincidences has nothing to do with these arguments on the
efficiency of the detectors because, as clearly explained in these arguments,
the efficiency is calibrated with the ratio of the number of particles
detected by the detector with polarizer and without polarizer, while
the core of the conditioning argument is that the number of detected
particles without polarizer may not be the total number of particles
emitted by the source. This number is usually unobservable in a real
experiment even if one postulates 100$\%$ efficiency in the detection.
In the real experiments (e.g. \cite{[WJSWZ98],SAWJST99,AsGrRo82}),
before taking the coincidences
we cannot speak of the total number of particles satisfying the singlet
condition.

Thus conditioning on coincidences has nothing to do with the loophole
argument. In this section we discuss the mathematical difference
between the two points of view and we show
how this can be experimentally verified. At the end
of this section we also explain the role of the dynamics (chameleon
effect) in establishing the coincidences.

In order to motivate and describe more precisely the difference between this
type of correlations and the unconditional ones, suppose we
have two particles $(1,2)$ with the same state space $S$ and let
\begin{equation}
(\sigma_{1,t})\ ;\quad(\sigma_{2,t})\ ;\quad(\sigma_t)=(\sigma_{1,t},
\sigma_{2,t})\ ;\quad t\in\Bbb R\label{stopr}
\end{equation}
be the stochastic processes describing the time evolution of these
particles in the state space (deterministic processes are a particular
case).

Suppose that the process (\ref{stopr}) is stationary and ergodic and
that we want to measure experimentally the equal time correlations of
two functions $S^{(1)}_a$, $S^{(2)}_b$ of these particles. 

If we do $N$ independent measurements in our laboratory at times
$t_1,\dots,t_N$ finding the results $\sigma_1,\dots,\sigma_N$ respectively,
can we conclude that
\begin{equation}
{1\over N}\,\sum^N_{j=1}S^{(1)}_a(\sigma_j)S^{(2)}_b(\sigma_j)\sim
E(S^{(1)}_a(t)S^{(2)}_b(t))=\langle S^{(1)}_a(t)S^{(2)}_b(t)\rangle
\label{corabs}
\end{equation}
as a naive application of the ergodic theorem would suggest? The answer
is clearly: no. In fact the states $\sigma_{1,t}$, $\sigma_{2,t}$ will
in general depend on many parameters such as position, momentum, spin,
polarization,...
$$\sigma_{1,t}=\sigma_{1,t}(q_{1,t},p_{1,t},\dots)$$
and, by our assumption, the effective measurements are done in the bounded
space regions $A_1$ and $B_2$. This means that we are not counting all the
emitted particles but only those which happened to be in the same time
in the regions $A_1$ and $B_2$ (recall that we are postulating 100\%
efficient detectors). This is precisely what characterizes
conditional probability. Therefore the correct conclusion is not
(\ref{corabs}) but
\begin{equation}\label{corcon}
{1\over N}\,\sum^N_{j=1}S^{(1)}_a(\sigma_j)S^{(2)}_b(\sigma_j)
\sim E\left(S^{(1)}_a(\sigma_t)S^{(2)}_b(\sigma_t) \ \mid \
q_1(t)\in A_1 \ ; \ q_2(t)\in B_2 \right)
\end{equation}
Thus, if we say that {\it a coincidence occurs\/} if, for some $t$ one
has both $q_1(t)\in A_1$ and $q_2(t)\in B_2$ and we denote
\begin{equation}
\Gamma_c(\sigma_t):=[q_1(t)\in A_1\quad\hbox{and}\quad q_2(t)\in B_2]
\label{coin}
\end{equation}
the corresponding event (notice that both $q_1(t)$ and $q_2(t)$ are
functions of $\sigma_t$), then we can rewrite (\ref{corcon}) in the form
\begin{equation}
{1\over N}\,\sum^N_{j=1}S^{(1)}_a(\sigma_j)S^{(2)}_b(\sigma_j)\sim
E(S^{(1)}_a(\sigma_{1,t})S^{(2)}_b(\sigma_{2,t})|\Gamma_c(\sigma_t)) =
\label{corcon2}
\end{equation}
$$={E(S^{(1)}_a(\sigma_t)S^{(2)}_b(\sigma_t)\chi_{\Gamma_c(\sigma(t))})
\over P(\Gamma_c(\sigma(t)))}$$
Because of the stationarity of the process, the right hand side of
(\ref{corcon2}) can be written
\begin{equation}
{E(S^{(1)}_aS^{(2)}_b\chi_{\Gamma_c})\over P(\Gamma_c)}\label{corcon3}
\end{equation}

The conclusion (\ref{corabs}) would be justified only if one could prove
that the number $N$, appearing in it, is the total number of emitted
pairs which, in many sitautions, is unboservable in principle.

As shown by the considerations above, the class of stochastic processes
such that there exist two regions $A_1$ and $B_2$ for which this is the
case, is very special (although surely nonempty).

In conclusion, let us consider a simple example, concerning a single
polarizer which may help clarifying the conceptual and experimental
difference between the efficiency and the coincidence problems.

Suppose that a detector is 100\% efficient. Then, if a source emits 100
photons, all photons are detected in absence of polarizer. Suppose
moreover that, when the polarizer is inserted, only 90 photons and not 100
are detected. Therefore, if as done in \cite{AsGrRo82}, the efficiency is
calibrated with the ratio of the number of particles detected by the detector
with polarizer and without polarizer, we should conclude that our polarizer
is 90\% efficient.

However, if the loss of these 10 photons is due to the chameleon
effect, then by repeating many times the experiment (and postulating a
situation of stationarity of the source) one should always detect 90
photons.

On the contrary, if the loss of photons is due to accidental
causes, then the number of detected photons should fluctuate and an
analysis of these fluctuations should, in principle, allow to
distinguish between an 100\% efficient detector in presence of the chameleon
effect and an 100\% efficient detector in presence of a 90\% efficient
polarizer.

In real physical situations the two effects are most likely combined
and their distinction, although clear in principle, might be a very
hard challenge both for theoreticians and experimentalists. However we
are convinced that a satisfactory theory of measurement should take into
account both these effects.

\section{Why contextuality is not enough}\label{context}

The following theorem shows that the contextuality argument alone, 
advocated by several authors\cite{Muynck,Khrennikov} is not
enough to rule out the application of Bell's inequality. In fact this
theorem implies that there is a large class of contextual hidden
variable theories, i.e. in which the initial distribution depends on the
global setting $a,b$ of the far away apparata, which satisfy the Bell
inequality and which therefore cannot reproduce the singlet
correlations. This shows in particular that to achieve such a violation

(i) dynamical considerations (chameleon effect) are necessary

(ii) in the local decomposition of our probability measure (\ref{locmeas})
the non normalization condition (\ref{nonorm}) of the local measures
$p_{1,a}, p_{2,b}$, i.e. the fact that they are not conditional probabilities 
in the sense of Remark (1) of Section (2), 
is a necessary condition for the above mentioned violation.

\begin{theorem}\label{imp1}
Let ${\cal A}_1,{\cal A}_2$ (system observables) and ${\cal A}_{M_1},
{\cal A}_{M_2}$ (apparatus observables) be commutative $*$--algebras
and, for any such algebra ${\cal A}$ denote ${\cal S}({\cal A})$ the set
of its states. Suppose that, for any pair of vectors in the unit sphere
$S^{(2)}$ in $3$--dimensional space:
\begin{description}
\item[(i)] the initial state of the system $(1,2,M_1,M_2)$ has the form
\begin{equation}\label{inst1}
\psi_{a,b}=\psi_{1,2}\circ(E_{1,a}\otimes E_{2,b})
\end{equation}
where $\psi_{1,2}\in{\cal S}({\cal A}_{1} \otimes {\cal A}_2)$ and
\begin{equation}\label{ce}
E_{j,x}:{\cal A}_j\otimes{\cal A}_{M_j}\to{\cal A}_j\quad  ;\quad
j=1,2;\ x\in S^{(2)}
\end{equation}
are conditional expectations.
\item[(ii)] For each $j=1,2$ and $x\in S^{(2)}$ it is given a self
adjoint element $S^{(j)}_x$ in ${\cal A}_j$ with spectrum in the
interval $[-1,1]$ and a positive normalized map $T_{j,x} :{\cal A}_j\to
{\cal A}_j$ ($T_{j,x}(1) =1$.
Then the pair correlations
\begin{equation}\label{corr1}
\langle S^{(1)}_aS^{(2)}_b\rangle=
\psi_{a,b}(T_{1,a}(S^{(1)}_a)T_{2,b}(S^{(2)}_b))=
\psi_{1,2}(E_{1,a}(T_{1,a}(S^{(1)}_a))
E_{2,b}(T_{2,b}(S^{(2)}_b)))
\end{equation}
cannot be the EPR ones.
\end{description}
\end{theorem}

\noindent{\it Proof\/}. By contradiction. Consider the 4 random
variables in ${\cal A}_{1}\otimes {\cal A}_{2}$:
\begin{equation}
E_{1,a}(T_{1,a}(S^{(1)}_a)),\ E_{2,b}(T_{2,b}(S^{(2)}_b)),E_{1,c}
(T_{1,c}(S^{(1)}_c)),E_{2,c}(T_{2,c}(S^{(2)}_c))\label{48a}
\end{equation}
The positivity and normalization of the dynamics implies that
\begin{equation}\label{vals}
-1\leq T_{j,x}(S^{(j)}_x)\leq+1 \quad  ;\quad j=1,2;\ x\in S^{(2)}
\end{equation}
hence the same inequalities hold for the random variables (\ref{48a}).
Moreover, if the correlations (\ref{corr1}) are equal to $-a\cdot b$
then, for any $c\in S^{(2)}$,
\begin{equation}
E_{1,c}(T_{1,c}(S^{(1)}_c))=-E_{2,c}(T_{2,c}(S^{(2)}_c))\ ;\quad
\psi_{1,2}-\hbox{a.e.}\label{sing}
\end{equation}
(cf. Lemma (14.1) in \cite{AcRe00b}) and from this the Bell's inequality
follows.

\begin{corollary}\label{imp2} If the initial state of the system has the
form
\begin{equation}
\psi_{12}\otimes\psi_{1,a}\otimes\psi_{2,b}\label{inst}
\end{equation}
with $\psi_{1,2}\in{\cal S}({\cal A}_{1}\otimes {\cal A}_{2})$,
$\psi_{j,x}\in{\cal S}({\cal A}_{M_j})$, $j=1,2$, $\forall\,x$ and if
the dynamics satisfy (\ref{vals}) then the pair correlations
\begin{equation}
\langle S^{(1)}_aS^{(2)}_b\rangle=(\psi_{12}\otimes\psi_{1,a}\otimes
\psi_{2,b})(T_{1,a}\otimes T_{2,b})(S^{(1)}_a\otimes
S^{(2)}_b)\label{corr}
\end{equation}
cannot be the EPR ones.
\end{corollary}

\noindent{\it Proof\/}. Apply Theorem (\ref{imp1}) to the case
$$E_{j,x}(a_j\otimes a_{M_1})=a_j\psi_{j,x}(a_{M_j})\ ;\quad j=1,2\ ;
\quad a_j\in{\cal A}_j\ ;\quad a_{M_j}\in{\cal A}_{M_j}$$

\section{ Conclusions}\label{concl}

The main point of the present experiment is not, as already emphasized
in the introduction, to build a hidden variable theory for the EPR
experiment: the main problem with hidden vaiable theories is not so much
their existence as their non uniqueness. This implies that an experimental
discrimination between any two such theories would be impossible due to
the Heisenberg principle.

The main idea of the "chameleon philosophy" can be summarized in the
words of A. Tartaglia \cite{[Tar98]}: "...
In conclusion quantum measurements and the story of the violation of the
Bell or similar inequalities tell us that the objects of the quantum world
are not like boxes containing spin, polarization vector etc. like buttons,
pins, pearls and the like, but like programmed machines capable of different
behaviours according to the physical conditions locally triggering them. 
..."

We have a great admiration for the extraordinary experiments that have
been done and continue to be done in this field and their value is in by no
way belittled by the present experiment.

Concerning the detection loophole argument we believe that the
distinction between chameleon effect (which has a principle nature) and
random errors (which are contingent) should stimulate a deeper
discussion of the very notion of "efficiency of an apparatus", leading
to a distinction, in the role played by the interaction system--apparatus
in the determination of the observed statistics, between the theoretically
describable, at least in principle, interaction between system and
apparatus and the accidental errors such as spurious photons,
imprecisions of clocks, imprecise determination of the initial state,
generic malfunctioning, ... . The existing models, based on the
efficiency argument, can be refined without much effort by including a
distinction between these two types of inefficiencies.

Finally, as far as quantum cryptography is concerned, we emphasize
the possible use of dynamical systems, such as the one discussed in the
present paper, both as a benchmark for truly quantum algorithms and as
a stimulus for potentially fruitful investigations on
hybrid algorithms, i.e. combining classical and quantum features, with a
potential impact on lowering the costs and increasing the performance of
the usual quantum cryptographic algorithms.

\section*{Acknowledgment}
The autors are grateful to Richard Gill whose fierce opposition to 
their ideas surely contributed to improve the presentation of the 
present paper.

\end{document}